# Laboratory Investigation of CO$_2$-Driven Enhancement of Radiolytic H$_2$O$_2$ on Europa and Other Icy Moons

Bereket D. Mamo,[1,2,3] Ujjwal Raut,[2,3,1] Ben D. Teolis,[2,3,1] Trevor P. Erwin,[4] Richard J. Cartwright,[5] Silvia Protopapa,[2,6] Kurt D. Retherford,[2,3,1] and Tom A. Nordheim[5]

[1] *Department of Physics and Astronomy, University of Texas at San Antonio, One UTSA Circle, San Antonio, TX 78249, USA*
[2] *Center for Laboratory Astrophysics and Space Science Experiments (CLASSE), Space Science Division, Southwest Research Institute, 6220 Culebra Road, San Antonio, TX 78238, USA*
[3] *Space Science Division, Southwest Research Institute, 6220 Culebra Road, San Antonio, TX 78238, USA*
[4] *Department of Earth, Atmospheric, and Planetary Sciences, Purdue University, 550 Stadium Mall Drive, West Lafayette, IN 47907*
[5] *Johns Hopkins University Applied Physics Laboratory, 11101 Johns Hopkins Rd, Laurel, MD 20723*
[6] *Solar System Science and Exploration Division, Southwest Research Institute, 1301 Walnut Street, Boulder, CO 80302*

## ABSTRACT

Observations of Europa's leading hemisphere reveal elevated H$_2$O$_2$ in the warmer, low latitude chaos terrains compared to the colder, polar regions. This distribution disagrees with prior laboratory radiolysis studies of pure water ice, which show higher H$_2$O$_2$ yields at colder temperatures. The regions with higher peroxide abundance, Tara and Powys Regiones, also present increased amounts of CO$_2$, possibly sourced from Europa's interior. To investigate whether CO$_2$ influences radiolysis of water ice to boost H$_2$O$_2$ production, we irradiated water ice doped with varying amounts of CO$_2$ with 10 keV electrons at 70 and 100 K. Our results indicate that CO$_2$, even in trace amounts ($< 3\%$), significantly enhances H$_2$O$_2$ yields at temperatures relevant to Europa. We discuss the mechanisms by which CO$_2$ boosts peroxide synthesis and quantify H$_2$O$_2$ creation and destruction cross sections and $G$-values across different CO$_2$ concentrations. These findings provide a plausible explanation for the perplexing H$_2$O$_2$ distribution on Europa and has implications for understanding peroxide on other icy bodies such as Ganymede and Charon, where it has been detected alongside CO$_2$.

## 1. INTRODUCTION

Hydrogen peroxide (H$_2$O$_2$), a radiolytic product of water ice, has been detected on the surfaces of icy bodies in the outer Solar System, including Europa (Carlson et al. 1999, 2009; Hand & Brown 2013), and more recently on Ganymede (Trumbo et al. 2023) and Charon (Protopapa et al. 2024). H$_2$O$_2$ forms primarily through the recombination of OH radicals, which are produced when water molecules are broken apart by energetic particles, including ions and electrons trapped in planetary magnetospheres, solar wind, and cosmic rays. Since its detection on Europa by the Galileo Near Infrared Mapping Spectrometer (Carlson et al. 1999), numerous laboratory studies have investigated radiolytic synthesis of H$_2$O$_2$ using energetic protons, heavy ions (Gomis et al. 2004; Loeffler et al. 2006; Moore & Hudson 2000), and electrons (Hand & Carlson 2011; Zheng et al. 2006a). These studies consistently demonstrate an inverse relationship between steady-state H$_2$O$_2$ abundance and ice temperature, with colder ice yielding more peroxide than warmer ice.

Surprisingly, Europa's peroxide distribution does not follow the temperature dependence predicted for pure water ice in the laboratory. Keck II and recent James Webb Space Telescope (JWST) data have revealed stronger H$_2$O$_2$ absorption bands in the warmer, low-latitude chaos terrains of the leading and anti-Jovian hemispheres, than in the colder, ice-rich high latitudes (Trumbo et al. 2019a; Wu et al. 2024). This $> 2\times$ enrichment of H$_2$O$_2$ in the warmer regions directly contradicts experimental findings, revealing a fundamental disagreement between laboratory predictions for pure water ice and Europa's peroxide distribution. A possible explanation is that regional compositional differences (chaos vs. polar regions) may exert a stronger influence in shaping Europa's H$_2$O$_2$ yields and distribution than ice

Corresponding author: Bereket D. Mamo
bereket.mamo@contractor.swri.org

temperature alone. The geologically young, disrupted chaos terrains, which potentially reflect recent interactions with the subsurface environment (e.g., Collins & Nimmo 2009; Schmidt et al. 2011), are notably enriched in $CO_2$ (Trumbo & Brown 2023; Villanueva et al. 2023). Could the presence of $CO_2$ drive the enhanced peroxide production in Europa's chaos regions, signaling a surface composition more conducive to the formation of this radiolytic oxidant? Supporting this hypothesis are preliminary experiments on irradiated $H_2O$–$CO_2$ ice mixtures that show increased $H_2O_2$ yields compared to pure water ice. Moore & Hudson (2000) attributed this enhancement to $CO_2$ molecules scavenging 'delta electrons'—secondary electrons generated through ionization of water molecules by primary projectiles. As electron acceptors, $CO_2$ molecules capture these secondary electrons, effectively shielding $H_2O_2$ from destruction by direct impact, dissociative electron attachment or other processes. While these earlier studies suggested a potential enhancement mechanism, they did not present a systematic analysis on how $CO_2$ amplifies the radiolytic synthesis of $H_2O_2$.

The correlation between Europa's $H_2O_2$ and $CO_2$ in JWST data (Raut et al. 2024; Wu et al. 2024) motivated us to conduct experiments to systematically investigate the effect of $CO_2$ inclusion on $H_2O_2$ synthesis. Using 10 keV electrons, we irradiated water ice containing varying amounts of $CO_2$ (up to $\sim 10\%$) and monitored the evolution of the 3.5 µm $H_2O_2$ absorption feature as a function of irradiation fluence (or equivalently dose). We find that small $CO_2$ concentrations ($\sim 3\%$) significantly enhance $H_2O_2$ production at 70 K, yielding a 2.5-fold increase compared to pure ice. Beyond 3%, this enhancement plateaus, with no noticeable increase in peroxide yield at higher $CO_2$ levels. At 100 K, the effect of $CO_2$ is even more drastic, with a 6× increase in $H_2O_2$ abundance. This temperature dependence suggests that diffusion-driven reactions between radiolytic products of $H_2O$ and $CO_2$ may contribute to the $H_2O_2$ enhancement. From the dose-response of peroxide synthesis in ice films with $CO_2$, we derive the $H_2O_2$ formation and destruction cross sections, the radiation yield ($G$-value), and steady-state abundance, and determine how these parameters vary with $CO_2$ concentration.

## 2. INSTRUMENTATION AND EXPERIMENTAL SETUP

The experimental setup has been described previously by Raut et al. (2022). Briefly, the experiments were performed in an ultra-high vacuum chamber (Figure 4A in Appendix A) that is cryo-pumped to a base pressure of $\sim 10^{-10}$ Torr. The chamber supports a 4 K compressed Helium cryostat that is mounted vertically on a rotatable stage. A 6 MHz quartz crystal microbalance (QCM) is embedded into a copper block bolted to the terminal end of the cryostat. Ice films were condensed onto the gold mirror electrode of the QCM, which is cooled to 70 K, using an omnidirectional/background vapor flux leaked from one or multiple gas dosers. The deposit causes a change in the QCM's resonant frequency in proportion to its areal mass (ng cm$^{-2}$), providing a measurement of the column density, $\eta$ (molecules cm$^{-2}$), if the deposit's molecular mass is known.

Films of pure $H_2O$ and $H_2O$ mixed with $CO_2$ were deposited at a rate of $\sim 40$ nm min$^{-1}$. At the temperature of ice deposition, we expect the phase to be amorphous (Mitchell et al. 2017; Sceats & Rice 1982). The column density of the films ranged from $\sim 7900$ to $\sim 8500$ monolayers (1 ML = $10^{15}$ molecules cm$^{-2}$), depending on the $CO_2$ concentration in the ice. This corresponds to a thickness of $\sim 2.73$ to $\sim 2.94$ µm for all films, calculated using densities $\rho_{H_2O} = 0.89$ g cm$^{-3}$ [from Fresnel equation fits to the UV-Visible reflectance spectra (Heavens 1991; Raut et al. 2008, 2007; Teolis et al. 2007)] and $\rho_{CO_2} = 1.6$ g cm$^{-3}$ (Loeffler et al. 2016; Satorre et al. 2008). This thickness is comparable to the projected range of the energetic electrons (Berger et al. 2017; Johnson 1990). For pure films, deposition was accomplished by condensing $H_2O$ vapor at a constant rate measured by the QCM. Mixed ice films with varying $CO_2$ concentrations were obtained by co-depositing from two dosers, one for $H_2O$ and another for $CO_2$ (Allodi et al. 2013). We initiated deposition by first introducing $CO_2$ gas and establishing a condensation rate (segment 1-2 in Figure 4B inset) on the QCM. Next, $H_2O$ was leaked in to increase the total rate to $\sim 40$ nm min$^{-1}$ (segment 2-3). At the end of deposition, the $H_2O$ doser was closed first to verify the terminal $CO_2$ deposition rate (segment 3-4) remained unchanged to within 10% of the initial value at the beginning of deposition. This simultaneous deposition technique allows us to precisely determine the abundance of $CO_2$, e.g., $(2.5 \pm 0.1)\%$ $CO_2$ by number in the icy mixture shown in Figure 4B in Appendix A. In some experiments, the uncertainties in $CO_2$ concentrations are larger ($\sim \pm 15\%$), owing to deviations or drifts from target deposition rates during co-deposition.

Following deposition, the pure $H_2O$ and $H_2O$–$CO_2$ mixed films were either kept at 70 K or warmed to 100 K prior to the onset of electron irradiation. The QCM records $CO_2$ loss (typically $< 40\%$ of the initial $CO_2$ amount at 70 K) that occurs during heating to 100 K; we correct for this decrease in the $CO_2$ abundance from the film due to thermal out-diffusion. The specular infrared reflectance spectra (1–16 µm) of the ice films were measured using a

Thermo-Nicolet is50 Fourier Transform Infrared spectrometer (FTIR) at an incidence angle of 35° and a resolution of 2 cm$^{-1}$. The spectra are expressed in optical depth as $-\ln(R/R_0)$, where $R$ and $R_0$ are the reflectance of the ice film and the gold substrate, respectively. The films were then irradiated at 70 and 100 K at normal incidence with a 10 keV electron beam from a Kimball Physics (EGG-3101) gun. The electron flux ($\sim 10^{12}$–$10^{13}$ e$^-$ cm$^{-2}$ s$^{-1}$) was measured with a Faraday cup placed adjacent to the QCM, and the beam was electrostatically rastered to ensure uniform irradiation of the films. Films were irradiated to a total fluence of $\sim 2.5 \times 10^{16}$ e$^-$ cm$^{-2}$ in incremental steps while acquiring IR spectra at each step.

## 3. RESULTS

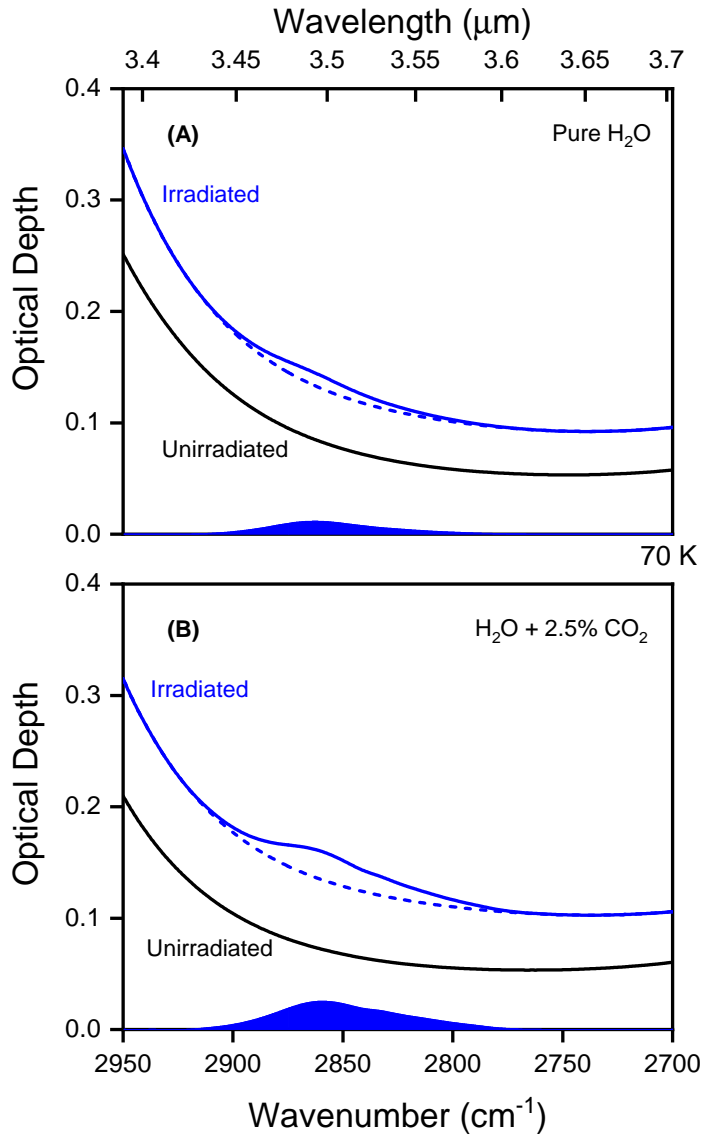

**Figure 1.** A: Emergence of the 3.5 μm $H_2O_2$ feature after irradiation of a 2.8 μm thick pure water ice film at 70 K with 10 keV electrons to a fluence of $1.5 \times 10^{16}$ e$^-$ cm$^{-2}$. A polynomial continuum (dashed blue curve) is subtracted from the irradiated spectra to obtain the asymmetric absorption (blue shaded region). B: Same as top panel but for water ice film of similar thickness that contains 2.5% $CO_2$. The integrated area of the continuum-subtracted absorption is $> 2.5\times$ larger for the $CO_2$-doped film relative to that of pure water ice. The spectra have been vertically shifted for clarity.

Figure 1A shows the emergence of the 3.5 μm absorption in the infrared spectrum of a pure water ice film irradiated with 10 keV electrons (solid blue curve) to a fluence of $1.5 \times 10^{16}$ e$^-$ cm$^{-2}$ (equivalent to a dose of 10 eV/molecule) at

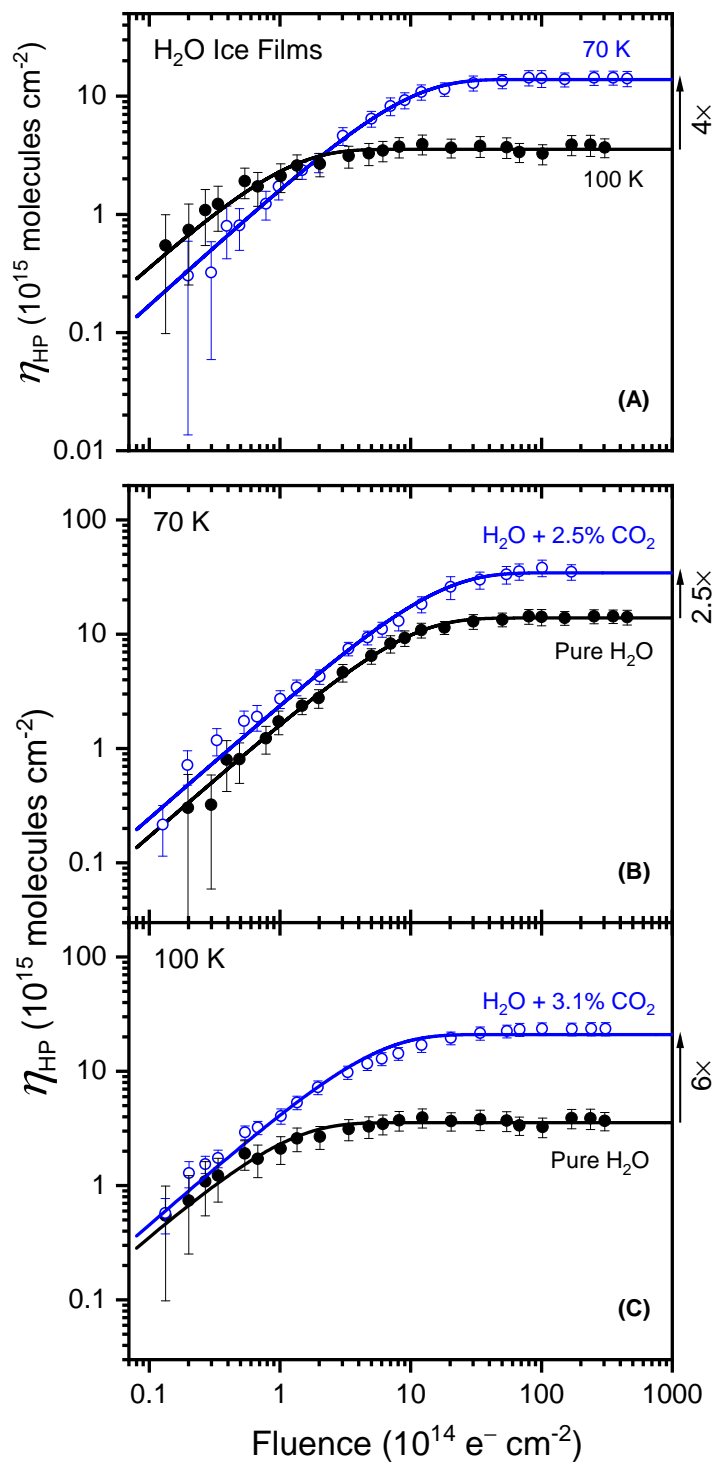

**Figure 2.** Fluence-dependent synthesis of $H_2O_2$ in 2.8 µm thick pure and $CO_2$-containing water ice films irradiated with 10 keV electrons at 70 and 100 K. Panel A: In pure water ice, $H_2O_2$ yields are higher at 100 K at low fluences but this trend reverses above $3 \times 10^{14}$ e$^-$ cm$^{-2}$, with the 70 K ice presenting nearly 4× higher steady-state peroxide abundance. Solid curves are model fit to the data discussed in the text. Panels B and C: Trace inclusions of $CO_2$ markedly increase $H_2O_2$ yields, with greater enhancements at warmer temperatures. At 70 K, $H_2O$ ice with 2.5% $CO_2$ shows a $\sim 2.5$-fold increase (indicated by the arrows), while at 100 K, ice with 3.1% CO2 yields a notable 6-fold enhancement over pure ice. Minor organics contributions to the 3.5 µm absorption have been removed. Mechanisms underlying this $CO_2$-induced boost in $H_2O_2$ are discussed in the text.

70 K. The asymmetric 3.5 µm absorption (blue shaded region) was isolated by subtracting a polynomial continuum (dotted blue curve) from the irradiated spectrum. The black curve is the unirradiated ice spectrum which guides the choice of the polynomial to reproduce the continua in the irradiated films. Following irradiation to the same dose, a film containing 2.5% $CO_2$ exhibits a $> 2.5\times$ increase in the integrated area of this continuum-removed absorption (Figure 1B).

The 3.5 µm absorption observed in electron-irradiated pure water ice films is produced by the $\nu_1 + \nu_6$ combination mode and/or the $2\nu_6$ overtone of the $H_2O_2$ molecule (Bain & Giguère 1955; Giguère & Harvey 1959; Miller & Hornig 1961). In $CO_2$-containing ices, this broad 3.5 µm band ($\sim 0.1$ µm width) is not exclusive to $H_2O_2$ but includes minor contributions from radiolytically produced CHO-organics such as methanol, which absorbs at 3.53 µm (Luna et al. 2018; Hodyss et al. 2009). This is consistent with a shoulder feature that appears on the long-wavelength side of the 3.5 µm band in the irradiated spectra of $CO_2$-doped ices (Figure 6A in Appendix C). We estimate and remove this organic contribution ($< 15\%$ of integrated band area) from the 3.5 µm band using spectral deconvolution (see Appendix C for details) to obtain a 'cleaned-up' $H_2O_2$ absorption. We examine next how this isolated $H_2O_2$ feature depends on ice temperature, irradiation dose, and $CO_2$ concentration.

Figure 2A shows the fluence dependence of $H_2O_2$ column density ($\eta_{HP}$) in pure water ice films irradiated with 10 keV electrons at 70 and 100 K. Column densities were calculated by dividing the organics-subtracted 3.5 µm band area by the absorption strength of $(4.7 \pm 0.5) \times 10^{-17}$ cm molecule$^{-1}$. This value, adapted from (Loeffler et al. 2006), has been adjusted for the thickness of our ice films, accounting for optical interference effects (Teolis et al. 2007). Initially, $H_2O_2$ abundance increases linearly with fluence ($F$) at both temperatures, though it remains consistently lower at 70 K. However, as irradiation continues, this temperature trend reverses. At 100 K, the $H_2O_2$ abundance levels off at $\sim 2 \times 10^{15}$ $H_2O_2$ cm$^{-2}$ following irradiation to a fluence of $\sim 10^{14}$ e$^-$ cm$^{-2}$. In contrast, at 70 K, the abundance continues to increase until it approaches a steady-state value of $\sim 10^{16}$ $H_2O_2$ cm$^{-2}$ at a higher fluence of $\sim 10^{15}$ e$^-$ cm$^{-2}$. Not only is the steady-state yield 4-times smaller in the 100 K ice, but it requires $10\times$ less fluence to reach that plateau, signaling more efficient peroxide destruction at warmer temperatures, likely related to the increased mobility of OH radicals. Our fluence dependence generally agrees with previous measurements [Figure 4 of (Hand & Carlson 2011)].

Figure 2B shows the effect of a small (2.5% concentration) $CO_2$ inclusion on peroxide synthesis at 70 K. The steady state fluence is similar between pure (black circles) and $CO_2$ (blue circles) doped ice, $\sim 10^{15}$ e$^-$ cm$^{-2}$. However, even this small amount of $CO_2$ significantly boosts the $H_2O_2$ yield at all fluence. At steady-state, the $H_2O_2$ yield more than doubles ($\sim 2.5$) compared to pure $H_2O$ ice. The enhancement is even more striking at warmer temperatures, with a similar 3.1% $CO_2$ mixture producing a factor of six increase in peroxide yield at 100 K compared to pure $H_2O$ ice at the same temperature (Figure 2C). To understand the mechanisms driving this remarkable $CO_2$-enhanced $H_2O_2$ synthesis, we irradiated water ice containing different $CO_2$ concentrations (up to 10%) and analyzed the fluence dependence to obtain peroxide $G$-values (defined as the number of molecules produced per 100 eV of absorbed energy) and creation and destruction cross sections, which we present next.

## 4. DISCUSSION

First, we consider the baseline case of radiolytic $H_2O_2$ synthesis in pure water ice, which begins with reactions between radicals generated in the ice by:

i) ionization of $H_2O$ molecules by the incoming projectile $X$ (10 keV e$^-$ in our experiments)

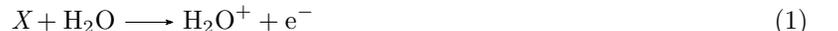
$$X + H_2O \longrightarrow H_2O^+ + e^- \qquad (1)$$

ii) electronic excitation of $H_2O$, followed by dissociation,

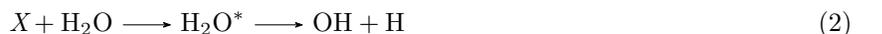
$$X + H_2O \longrightarrow H_2O^* \longrightarrow OH + H \qquad (2)$$

iii) and dissociative electron attachment (DEA) of low-energy secondary electrons [generated in (1)] to water molecules to form a transient anion that dissociates into a negative OH and a neutral hydrogen atom

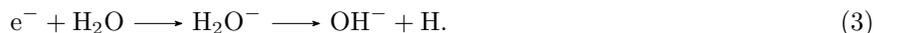
$$e^- + H_2O \longrightarrow H_2O^- \longrightarrow OH^- + H. \qquad (3)$$

Subsequent combination of OH radicals leads to the formation of $H_2O_2$ in the ice

$$2\,\mathrm{OH} \longrightarrow H_2O_2. \tag{4}$$

$H_2O$ can also dissociate into $H_2$ and $O(^1D)$, though gas-phase studies (Slanger & Black 1982; Okabe 1978) indicate this pathway occurs with only 10% yield. In the solid phase, the hot $O(^1D)$ atoms may react with surrounding water molecules to form $H_2O_2$ (Zheng et al. 2006b,a), or quench to the ground state via collisions and/or recombine with $H_2$ to reform water (Schriever et al. 1991; Tarasova et al. 1993; Johnson & Quickenden 1997; Teolis et al. 2017). As shown in Figure 2A, $H_2O_2$ abundance initially increases monotonically with fluence because, at small concentrations, its destruction is negligible. However, as the $H_2O_2$ level rises, so too does the rate of destruction by incoming projectiles or secondary electrons, (5), or radicals in the ice, (6) and (7)

$$e^- + H_2O_2 \longrightarrow H_2O + O^-, \tag{5}$$
$$\mathrm{OH} + H_2O_2 \longrightarrow H_2O + HO_2, \tag{6}$$
$$\mathrm{H} + H_2O_2 \longrightarrow \mathrm{OH} + H_2O. \tag{7}$$

Eventually at high fluence, the rates of $H_2O_2$ formation and destruction come into balance, causing the abundance to plateau. A simple model describing the peroxide fluence dependence incorporates two competing processes—$H_2O_2$ creation and destruction—each governed by its own cross section, $\sigma_{c_0}$ and $\sigma_{d_0}$, respectively,

$$\frac{d\eta_{HP}}{dF} = \sigma_{c_0}\eta_w - \sigma_{d_0}\eta_{HP}, \tag{8}$$

where $\eta_w$ and $\eta_{HP}$ are the column densities (in molecules cm$^{-2}$) of water and peroxide and $F$ is the electron fluence. Neglecting the minor change in $\eta_w$ during irradiation, eq. (8) integrates to

$$\eta_{HP}(F) = \eta_{HP_\infty}[1 - \exp(-\sigma_{d_0}F)], \tag{9}$$

The quantity $\eta_{HP_\infty} = \eta_w \times (\sigma_{c_0}/\sigma_{d_0})$ is the steady-state $H_2O_2$ column density at terminal $F$. We fit eq. (9) to the $\eta_{HP}$ vs. $F$ data for pure $H_2O$ ice at 70 and 100 K (Figure 2A) to obtain $\sigma_{c_0}$ and $\sigma_{d_0}$ and derive a $G$-value from the relation $G(\text{molecules}/100\,\text{eV}) = \eta_{HP_\infty} \times (\sigma_{d_0}/E) \times 100$, where $E$ is the incident electron energy (Teolis et al. 2017). Equivalently, the $G$-value can also be derived from the slope of the $H_2O_2$ abundance vs. $F$ in the low fluence regime, where destruction is negligible (Figure 2A).

At low fluence ($< 10^{14}$ e$^-$ cm$^{-2}$), the $H_2O_2$ abundance at 100 K actually exceeds that at 70 K (Figure 2A) likely due to enhanced thermal mobilization of OH radicals. Around 100 K, more OH radicals are thermally released from trapping sites in the ice lattice (Bednarek & Plonka 1987; Johnson & Quickenden 1997; Quickenden et al. 1991; Siegel et al. 1960), and therefore, can readily combine with other OHs to form peroxide. This enhanced recombination is reflected in the greater $\sigma_{c_0}$ and $G$-value observed at 100 K compared to 70 K (Figures 3A and 3B; data points for pure water ice with no $CO_2$). However, as $H_2O_2$ abundance increases with $F$, the probability of OH radicals encountering and destroying $H_2O_2$ by reaction (6) also rises at higher temperature. This is reflected in $\sigma_{d_0}$ at 100 K, which surpasses the 70 K value by about an order of magnitude (Figure 3C).

Next, we discuss the effects of $CO_2$ on $H_2O_2$ production in mixtures. The addition of $CO_2$, even in trace amounts, introduces a plethora of new reaction channels that collectively increase $H_2O_2$ synthesis in the ice. The most direct pathway is the release of energetic O atoms from $CO_2$, which can directly react with nearby $H_2O$ molecules, contributing to $H_2O_2$ creation. Another pathway involves H abstraction by CO which may lead to OH enrichment in the ice. During irradiation, $CO_2$ molecules dissociate into O and CO, and CO can be hydrogenated in multiple steps to form organics, such as HCO (1855 cm$^{-1}$), $H_2CO$ (1498 cm$^{-1}$), and $CH_3OH$ (2831 and 1016 cm$^{-1}$), identified in the our IR spectra (Figure 5 in Appendix B), in agreement with (Gerakines et al. 2000; Luna et al. 2018; Schutte et al. 1993). These 'organic forming' reactions deplete the ice of H atoms that would otherwise recombine with OH to form water. This results in excess OHs in the ice to create more $H_2O_2$.

The inclusion of $CO_2$ can also enhance $H_2O_2$ abundance by inhibiting its destruction. This occurs as $CO_2$ or its radiolytic by-products react with OH, thereby inhibiting $H_2O_2$ breakdown by reaction (6). Furthermore, the electron scavenging property of $CO_2$ (Moore & Hudson 2000) offers a protective mechanism that preserve $H_2O_2$ in the ice against secondary electrons that would otherwise destroy $H_2O_2$ by DEA or, given sufficient energy, by direct impact

$$e^- + H_2O_2 \longrightarrow \mathrm{OH}^- + \mathrm{OH}, \tag{10}$$

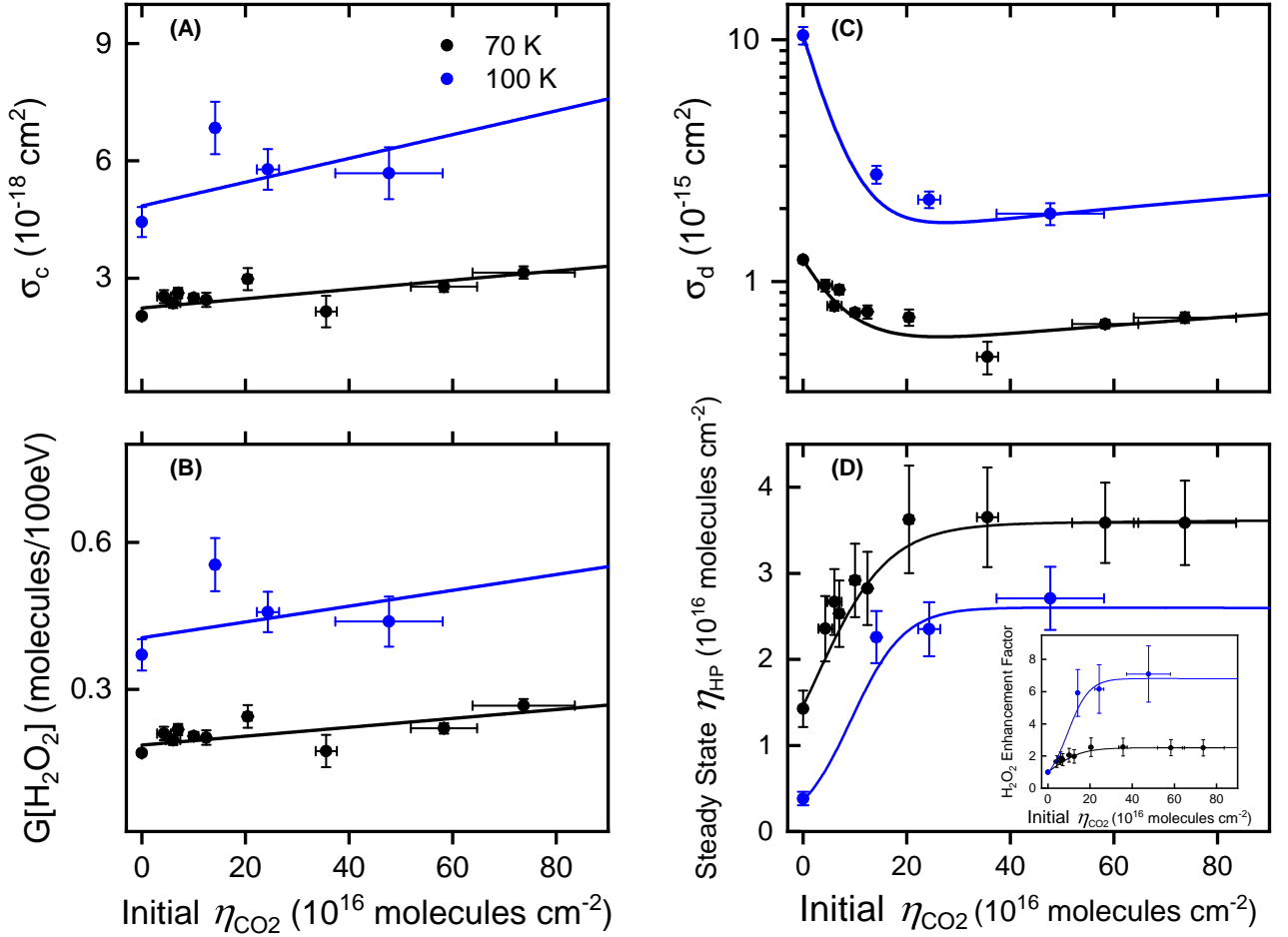

**Figure 3.** Dependence of the $H_2O_2$ creation cross section ($\sigma_c$), $G$-value, $H_2O_2$ destruction cross section ($\sigma_d$), and steady-state $H_2O_2$ abundance on the initial $CO_2$ abundance in water ice irradiated at 70 and 100 K. Both $\sigma_c$ and $G$-value increase linearly with $CO_2$ amount, with steeper slopes at 100 K (Panels A and B). This rise suggests that $CO_2$ promotes peroxide synthesis primarily via multiple H-atom abstraction by CO, synthesizing CHO-organics and resulting in an excess of OH radicals. More striking is the sharp decline in $\sigma_d$ with $CO_2$ inclusion (Panel C), which has a steeper plunge at 100 K ($\sim 5x$) than at 70 K ($\sim 3x$). This sharp decrease in $\sigma_d$ is likely due to $CO_2$-mediated shielding mechanisms that protect $H_2O_2$ from destruction by either secondary electrons or OH radicals. The scavenging of electrons and/or OH radicals by $CO_2$ or its radiolytic products is expected to be more efficient at higher temperatures, hence the steeper decline at 100 K. A small rise in $\sigma_d$ at larger $CO_2$ levels reflects $H_2O_2$ destruction facilitated by $CO_2$. Panel D shows the steady-state $H_2O_2$ abundance, $(\sigma_c/\sigma_d) \times \eta_w$, plotted against $CO_2$ amount. The sharp rise in $H_2O_2$ yield mirrors the decline in $\sigma_d$ below 3% $CO_2$. See text for further discussion on mechanisms underlying these results.

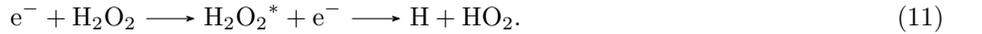

$$e^- + H_2O_2 \longrightarrow H_2O_2^* + e^- \longrightarrow H + HO_2. \tag{11}$$

By capturing these delta electrons into their orbitals and forming a transient $CO_2^-$ anion, $CO_2$ may shield $H_2O_2$ from direct destruction by channels (10) and (11). Therefore, $CO_2$ increases $H_2O_2$ synthesis by (i) direct O addition to $H_2O$ (ii) boosting OH population and (iii) shielding peroxide molecules from destruction by OH radicals and secondary electrons. These processes collectively elevate the $H_2O_2$ abundance in $CO_2$-doped ices relative to pure $H_2O$ ice. However, at higher concentrations, $CO_2$ or its radiolytic products may also react directly with $H_2O_2$, destroying it to form other organics (Moore & Hudson 2000; Pilling et al. 2010). We incorporate these mechanisms by modifying eq. (8) to include $CO_2$-induced creation, destruction, and shielding processes regulated by cross sections, $\sigma_{c_1}$, $\sigma_{d_1}$, and $\sigma_s$, respectively, all of which are assumed to be functions of the $CO_2$ column density $\eta_{CO_2}$:

$$\begin{aligned}\frac{d\eta_{HP}}{dF} &= \sigma_{c_0}\eta_w + \sigma_{c_1}(\eta_{CO_2})\eta_w \frac{\eta_{CO_2}}{\eta_T} - \sigma_{d_0}\eta_{HP} - \sigma_{d_1}(\eta_{CO_2})\eta_{HP}\frac{\eta_{CO_2}}{\eta_T} + \sigma_s(\eta_{CO_2})\eta_{HP} \\ &= \sigma_c(\eta_{CO_2})\eta_w - \sigma_d(\eta_{CO_2})\eta_{HP}.\end{aligned} \tag{12}$$

Here $\eta_T$ is the total column density, which is approximated as the sum of water ($\eta_w$) and $CO_2$ ($\eta_{CO_2}$). Neglecting minor changes in $\eta_w$ and $\eta_{CO_2}$ ($\sim 75\%$ of the initial $CO_2$ survives irradiation; Figure 7A in Appendix D), eq. (12) integrates to

$$\eta_{HP}(F) = \eta^*_{HP_\infty}[1 - \exp(-\sigma_d F)], \qquad (13)$$

where $\eta^*_{HP_\infty} = \eta_w \times (\sigma_c/\sigma_d)$ is the steady-state $H_2O_2$ column density in $CO_2$-doped ices and $\sigma_c$ and $\sigma_d$ are the $CO_2$ concentration-dependent *effective* creation and destruction cross sections, given by $\sigma_c = \sigma_{c_0} + \sigma_{c_1}(\eta_{CO_2}) \times (\eta_{CO_2}/\eta_T)$ and $\sigma_d = \sigma_{d_0} - \sigma_s(\eta_{CO_2}) + \sigma_{d_1}(\eta_{CO_2}) \times (\eta_{CO_2}/\eta_T)$. We fit eq. (13) to the $H_2O_2$ column density vs. $F$ data for $CO_2$–$H_2O$ icy mixtures (Figures 2B and 2C) to obtain $\sigma_c$, $\sigma_d$, and $G$-values presented in Figure 3.

Up to the maximum measured $CO_2$ concentration of $\sim 10\%$ (Table 1), the creation cross section $\sigma_c$ and the $G$-value scale linearly with $CO_2$ abundance at both 70 and 100 K (Figures 3A and 3B). The slope of $\sigma_c$ (i.e. $\sigma_{c_1}$) and that of the $G$-value are $\sim 2.5\times$ and $\sim 2\times$ higher at 100 K than at 70 K, respectively, consistent with (i) increased availability of OH radicals due to efficient H sequestration by CO to form organics and (ii) enhanced OH mobility at higher temperature. This interpretation is supported by the higher abundance of carbonic acid ($H_2CO_3$) at 100 K than at 70 K that desorbs from the microbalance at $\sim 255$ K (Figure 7 in Appendix D).

Remarkably, adding trace amounts of $CO_2$ (up to 3% concentration) sharply reduces $\sigma_d$, with a stronger decline at 100 K ($\sim 5\times$ reduction relative to pure ice case) than at 70 K ($\sim 2.5\times$; Figure 3C). This decline is consistent with (i) $CO_2$ shielding $H_2O_2$ from destruction by secondary electrons and (ii) radiolytic products of $CO_2$ (CO and O) reacting with OH radicals that would otherwise break down $H_2O_2$. Both electron scavenging and CO/O reactions with OH radicals are diffusion-dependent, and thus expected to be more efficient at elevated temperatures. Hence, the larger decline in $\sigma_d$ in the warmer 100 K ice compared to 70 K. Furthermore, electron scavenging by $CO_2$ molecules leads to formation of transient $CO_2^-$ anion, which can readily react with H to form organics. This reaction again sequesters H into CHO organics contributing to further enrichment of OH radicals favorable to $H_2O_2$ synthesis. We note $\sigma_d$ shows a small, but steady increase at $CO_2$ concentrations $> 7\%$.

The non-linear decline of $\sigma_d$ with $CO_2$ abundance is fitted with an empirical relation (Figure 3C), $\sigma_d = \sigma_{d_0} - \sigma_{s_1}(1 - \exp(-\sigma_{s_0}\eta_{CO_2})) + \sigma_{d_1} \times (\eta_{CO_2}/\eta_T)$, where the second term captures the protective effect of $CO_2$ against secondary electrons and OH radicals, which lowers the destruction cross section relative to pure ice value ($\sigma_{d_0}$ term in equation). The third linear term accounts for the $CO_2$-induced $H_2O_2$ destruction. An example of this empirical fit and the associated parameters is shown in Figure 8 of Appendix E. Note that this empirical relation may not be valid beyond the $CO_2$ concentrations investigated here ($< 10\%$). The cross sections for pure ($\sigma_{c_0}$, $\sigma_{d_0}$) and $CO_2$-mixed ($\sigma_c$, $\sigma_d$) water ice, $G$-values, and steady-state $H_2O_2$ concentrations are provided in Table 1, alongside values from previous electron irradiation studies of pure water ice.

The $H_2O_2$ synthesis and destruction rates are balanced at large $F$ resulting in steady state yields. We plot in Figure 3D the measured steady-state abundance of radiolytic $H_2O_2$ vs. initial $CO_2$ column density prior to irradiation at 70 and 100 K. At both temperatures, the steady state peroxide abundance rises sharply, but plateaus above $\sim 3\%$ $CO_2$ concentration. The sharp increase in $H_2O_2$ yield at small $CO_2$ abundances mirrors the marked decrease in $\sigma_d$ shown in Figure 3C, since

$$\eta^*_{HP_\infty} = \eta_w \times (\frac{\sigma_c}{\sigma_d}) = \eta_w \times \frac{\sigma_{c_0} + \sigma_{c_1}\frac{\eta_{CO_2}}{\eta_T}}{\sigma_{d_0} - \sigma_{s_1}(1 - \exp(-\sigma_{s_0}\eta_{CO_2})) + \sigma_{d_1}\frac{\eta_{CO_2}}{\eta_T}}. \qquad (14)$$

The steady-state $H_2O_2$ vs. initial $CO_2$ abundance data in Figure 3D are fitted by taking the ratio of the fits for $\sigma_c$ (Figure 3A) and $\sigma_d$ (Figure 3C), each plotted against initial $\eta_{CO_2}$. The inset shows that adding the same $\sim 3\%$ $CO_2$ concentration enhances $H_2O_2$ formation more strongly at warmer temperatures; at 100 K, it leads to a 6-fold increase in $H_2O_2$ abundance, whereas the same $CO_2$ inclusion results in only a 2.5-fold increase at 70 K. Beyond 3%, the enhancement in $H_2O_2$ yield appears to plateau, at least up to 10% $CO_2$. Adding $CO_2$ at higher concentrations results in declining $H_2O_2$ yields, as indicated in experiments by Pilling et al. (2010), although Strazzulla et al. (2005) find increased yields at all concentrations investigated. It is not clear if contribution of organics to the 3.5 µm band is accounted for in both studies.

## 5. CONCLUSION AND ASTROPHYSICAL IMPLICATIONS

The discovery of drastically boosted hydrogen peroxide production, obtained by doping electron-irradiated water ice with only miniscule (3%) concentrations of $CO_2$, may be a pivotal clue to understanding Europa's anomalous surface peroxide distribution. The principal mechanism is an unexpected major reduction in the $H_2O_2$ destruction cross

**Table 1.** Key experimental parameters, derived *G*-values, cross sections, and steady state concentration, alongside comparison to previous electron irradiation studies.

| Data Set | Energy (keV) | Ice Temp (K) | $\eta_w$ ($10^{18}$ $H_2O$ cm$^{-2}$) | $\eta_{CO_2}$ ($10^{16}$ $CO_2$ cm$^{-2}$) | $CO_2$ Concentration (%) | $\sigma_c$ ($10^{-16}$ cm$^2$) | $\sigma_d$ ($10^{-16}$ cm$^2$) | $G$ (molecules/100eV) | Steady-state $H_2O_2$ (%) |
|---|---|---|---|---|---|---|---|---|---|
| This work | 10 | 70 | 8.4 | 0.0 | 0.0 | $0.0203 \pm 0.0007$ | $12.3 \pm 0.4$ | $0.170 \pm 0.006$ | $0.17 \pm 0.02$ |
| | | | 8.3 | 4.3 | 0.5 | $0.025 \pm 0.002$ | $9.6 \pm 0.5$ | $0.21 \pm 0.01$ | $0.26 \pm 0.04$ |
| | | | 8.2 | 6.0 | 0.7 | $0.024 \pm 0.001$ | $7.9 \pm 0.4$ | $0.20 \pm 0.01$ | $0.30 \pm 0.04$ |
| | | | 8.2 | 7.0 | 0.8 | $0.026 \pm 0.001$ | $9.2 \pm 0.4$ | $0.22 \pm 0.01$ | $0.28 \pm 0.04$ |
| | | | 8.1 | 10.0 | 1.2 | $0.025 \pm 0.001$ | $7.4 \pm 0.3$ | $0.205 \pm 0.008$ | $0.34 \pm 0.05$ |
| | | | 8.1 | 12.4 | 1.5 | $0.024 \pm 0.002$ | $7.5 \pm 0.5$ | $0.20 \pm 0.01$ | $0.33 \pm 0.05$ |
| | | | 8.0 | 20.4 | 2.5 | $0.030 \pm 0.003$ | $7.1 \pm 0.5$ | $0.24 \pm 0.02$ | $0.42 \pm 0.07$ |
| | | | 7.8 | 35.6 | 4.4 | $0.021 \pm 0.004$ | $4.9 \pm 0.8$ | $0.17 \pm 0.03$ | $0.44 \pm 0.07$ |
| | | | 7.3 | 58.3 | 7.4 | $0.028 \pm 0.001$ | $6.7 \pm 0.3$ | $0.22 \pm 0.01$ | $0.42 \pm 0.05$ |
| | | | 7.7 | 73.7 | 8.7 | $0.031 \pm 0.002$ | $7.1 \pm 0.4$ | $0.27 \pm 0.01$ | $0.44 \pm 0.06$ |
| | | 100 | 8.4 | 0.0 | 0.0 | $0.044 \pm 0.004$ | $104 \pm 9$ | $0.37 \pm 0.03$ | $0.043 \pm 0.008$ |
| | | | 8.0 | 14.1 | 1.7 | $0.068 \pm 0.007$ | $28 \pm 2$ | $0.55 \pm 0.05$ | $0.25 \pm 0.03$ |
| | | | 7.7 | 24.3 | 3.1 | $0.058 \pm 0.005$ | $22 \pm 2$ | $0.46 \pm 0.04$ | $0.26 \pm 0.03$ |
| | | | 7.2 | 47.7 | 6.2 | $0.057 \pm 0.007$ | $19 \pm 2$ | $0.44 \pm 0.05$ | $0.30 \pm 0.04$ |
| Hand & Carlson (2011) | 10 | 80 | $9^a$ | 0.0 | 0.0 | $0.004^b$ | 8.7 | $0.028 \pm 0.003$ | $0.043 \pm 0.002$ |
| | | 100 | $9^a$ | 0.0 | 0.0 | $0.007^b$ | 38.5 | $0.050 \pm 0.004$ | $0.029 \pm 0.001$ |
| | | 120 | $9^a$ | 0.0 | 0.0 | $0.001^b$ | 18.2 | $0.009 \pm 0.001$ | $0.0063 \pm 0.0006$ |
| Zheng et al. (2006a) | 5 | 60 | 0.3 | 0.0 | 0.0 | 0.0065 | 1.5 | $\cdots$ | 0.4 |
| | | 90 | 0.3 | 0.0 | 0.0 | $\cdots$ | $\cdots$ | $\cdots$ | 0.03 |

NOTE—For pure ice $\sigma_c = \sigma_{c_0}$ and $\sigma_d = \sigma_{d_0}$. The parameters $\eta_w$, $\eta_{CO_2}$, and $CO_2$ Concentration represent initial values before irradiation.

[a] $\eta_w$ determined by equating film thickness to projected range of 10 keV electrons [see Hand & Carlson (2011)]

[b] Corrected for typographical error in exponent [cf. Hand & Carlson (2011)]

section, by a factor of $\sim 5\times$ with only $\sim 3\%$ $CO_2$, as $CO_2$ shields $H_2O_2$ from destruction by scavenging destructive secondary electrons and OH radicals. Additionally, $CO_2$ promotes peroxide synthesis by enhancing OH availability in the ice, achieved through multiple H atom abstraction to form CHO-organics (Figure 5 in Appendix B), as well as through direct hot O atom addition to $H_2O$ molecules. These findings significantly advance the understanding of peroxide synthesis on icy moons, especially the role of trace surface constituents like $CO_2$, crucial for interpreting $H_2O_2$ surface abundance and distribution.

Our experiments confirm a strong correlation between $H_2O_2$ and $CO_2$ in irradiated ice, which is broadly consistent with that found at Europa (although work is still ongoing to constrain Europa's absolute $CO_2$ concentrations). The decrease in $\sigma_d$, due to quenching of $H_2O_2$ destruction by secondary electrons and OH radicals, suggests that peroxide molecules in an $H_2O + CO_2$ matrix on Europa could be less susceptible to irradiation by magnetospheric particles than those in a pure ice matrix. Using the destruction cross section for pure $H_2O$ ice at 100 K (Table 1), the destruction rate per $H_2O_2$ molecule is $\sigma_d F = 1.9 \times 10^{-6}$ s$^{-1}$ for an electron flux at Europa $F = 1.8 \times 10^8$ cm$^{-2}$ s$^{-1}$ (Cooper et al. 2001). This rate corresponds to a characteristic time of $\sim$one week. Inclusion of 3% $CO_2$ extends this time to $\sim$one month.

The extended lifetime and increased $H_2O_2$ yields in $CO_2$-mixed ice, as observed in Tara and Powys Regiones, may have important implications for the habitability of Europa's subsurface ocean. Ocean-sourced $CO_2$ that enhances $H_2O_2$ production in the chaos terrains, could thereby boost the oxidant supply to Europa's underlying ocean, a key factor in determining its astrobiological potential. It is important to note, however, that the impact of endogenic brines such as NaCl, which is also enriched in the chaos terrains (Trumbo et al. 2019b, 2022), on peroxide production and destruction rates remains unclear and further experiments are needed to investigate their effect on oxidant yields.

The physical state of $CO_2$ on Europa's surface could also affect peroxide production in chaos terrains. While our laboratory experiments demonstrate that $CO_2$ intimately mixed with water ice enhances $H_2O_2$ yields, Europa's $CO_2$ likely exists in multiple physical states. JWST spectra of Tara Regio reveal a double-lobed absorption with peaks at 4.25 and 4.27 μm (Trumbo & Brown 2023; Villanueva et al. 2023). The 4.27 μm peak matches laboratory measurements of pure $CO_2$ ice. However, pure $CO_2$ ice is thermally unstable at the $\sim$100 K surface temperature of equatorial chaos terrains. Our spectra of $CO_2$ intimately mixed with water ice (Figure 5) show an absorption at 4.268 µm, consistent with previous studies (Moore & Hudson 2000; Gerakines et al. 2000; Pilling et al. 2010), though slightly blue-shifted (0.005 µm) compared to Gálvez et al. (2008). This suggests that a fraction of Europa's $CO_2$ could be intimately mixed within the water ice matrix, enabling thermal stability at higher temperatures. This form of $CO_2$, according to our experiments, could increase the radiolytic $H_2O_2$ yields in the chaos terrains.

The 4.25 µm lobe remains poorly understood but has been loosely ascribed to $CO_2$ adsorbed onto or trapped within brines such as NaCl, or to $CO_2$ embedded in a water-methanol matrix (Villanueva et al. 2023). However, methanol or other organics have not been detected in Europa's JWST spectra (Trumbo & Brown 2023; Villanueva et al. 2023). Systematic laboratory experiments investigating $CO_2$ in candidate briny and organics matrix are needed to clarify the origin of the 4.25 µm absorption, identify the host material, and assess whether $CO_2$ in such matrices could enhance peroxide production in the chaos terrains.

A similar correlation of radiolytic $H_2O_2$ to $CO_2$ might exist on other irradiated icy bodies. On Charon, JWST observations reveal that $CO_2$ exists in pure crystalline form and possibly in an intimately mixed state with $H_2O$ ice, alongside a distinct 3.5 µm $H_2O_2$ absorption feature that closely matches laboratory spectra of irradiated $CO_2$-containing water ice (Protopapa et al. 2024). The abundance of the intimately mixed state remains uncertain. However, if its presence is confirmed, particularly at shallower penetration depths, and its concentration is similar to that of the segregated component (estimated at $\sim 2\%$), then this would be sufficient, according to our experiments, to significantly enhance $H_2O_2$ formation. Likewise, Ganymede's polar regions, exposed to intense radiation, show peroxide in water ice-rich surfaces containing solid $CO_2$ (Bockelée-Morvan et al. 2024; Trumbo et al. 2023), suggesting that $CO_2$ could influence peroxide yields and distributions on this body as well. More broadly, our experimental findings suggest that only small deviations in surface composition at these bodies away from pure water ice may substantially enhance the synthesis of radiolytic oxidants such as peroxide in the surfaces.

## 6. ACKNOWLEDGMENTS

This work is primarily supported by NASA FINESST grant 80NSSC23K1386. The authors also acknowledge funding from NASA Solar System Workings grant 80NSSC24K0657 and support from the Europa Clipper Project 80NM0018D0004. Trevor Erwin was supported by the Europa ICONS (Inspiring Clipper: Opportunities for Next-generation Scientists) internship program. We also thank Hope Jerris, an ICONS intern, for assistance in electron beam alignment. Both ICONS interns were funded under Research Support Agreement #1711180. We thank M. J. Loeffler for valuable discussions and the two anonymous reviewers for helpful comments and suggestions.

# APPENDIX

## A. CHAMBER SCHEMATIC AND MICROBALANCE RESPONSE DURING ICE DEPOSITION

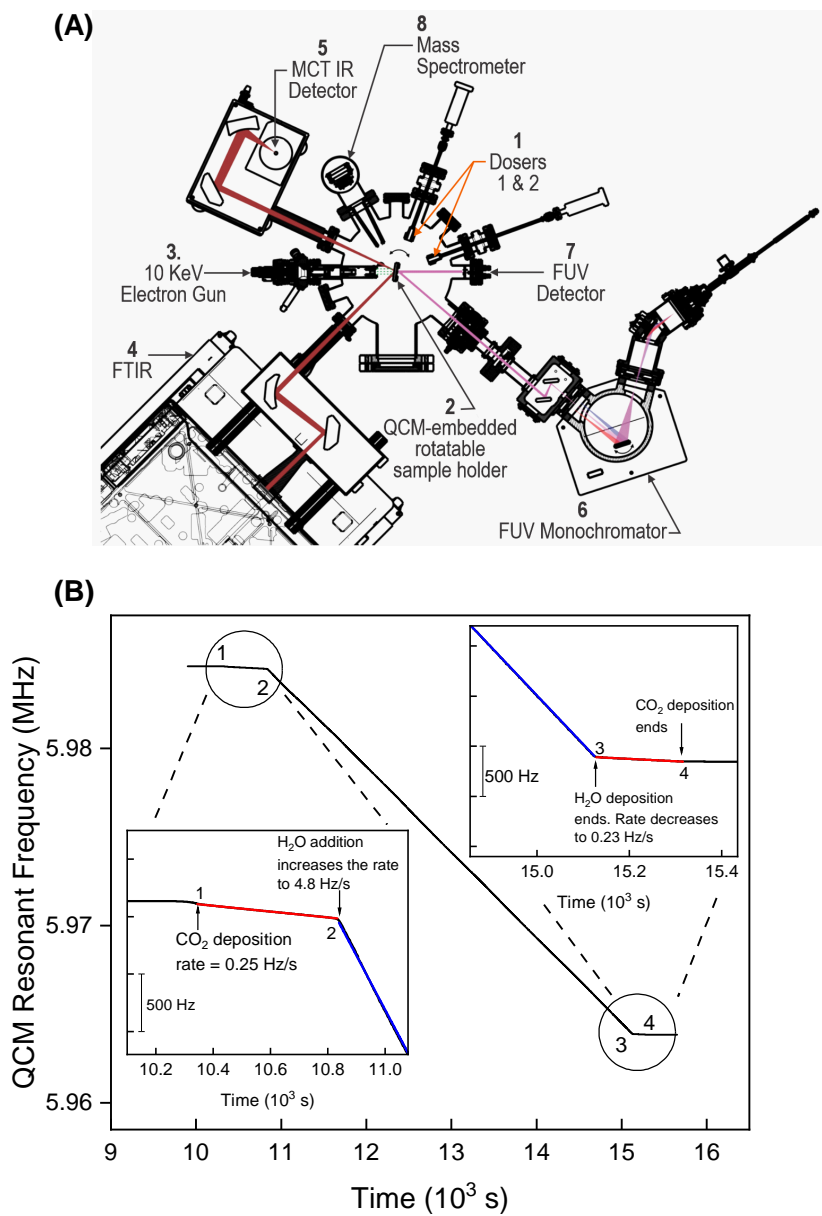

**Figure 4. A:** Schematic of the ultra-high vacuum chamber used for electron irradiation experiments of Europan ice analogs. **B:** Shift in the resonant frequency of the QCM during co-deposition of $H_2O$ and $CO_2$ ice from dedicated gas dosers. We first condense $CO_2$ [1] from a doser at a constant rate of $(0.25 \pm 0.01)$ Hz/s (zoomed in the inset). We then admit water vapor [2] through a second doser which increases the deposition rate to $(4.8 \pm 0.2)$ Hz/s. When the film approaches the desired thickness (2.8 μm, which corresponds to a frequency shift of $\sim 21,000$ Hz), we terminate $H_2O$ gas deposition first [3] and continue with $CO_2$ condensation briefly to ensure the terminal $CO_2$ rate did not drift significantly from the initial value ($< 10\%$). The $CO_2$ doser is shut off at [4]. This co-deposition scheme using multiple gas dosers with QCM gravimetry allows for producing mixed ices with precisely tailored composition. This film has $CO_2$ intimately mixed with water ice at $(2.5 \pm 0.1)\%$ (by number) concentration.

## B. RADIATION-INDUCED SPECTRAL CHANGES IN $H_2O$–$CO_2$ ICE MIXTURE

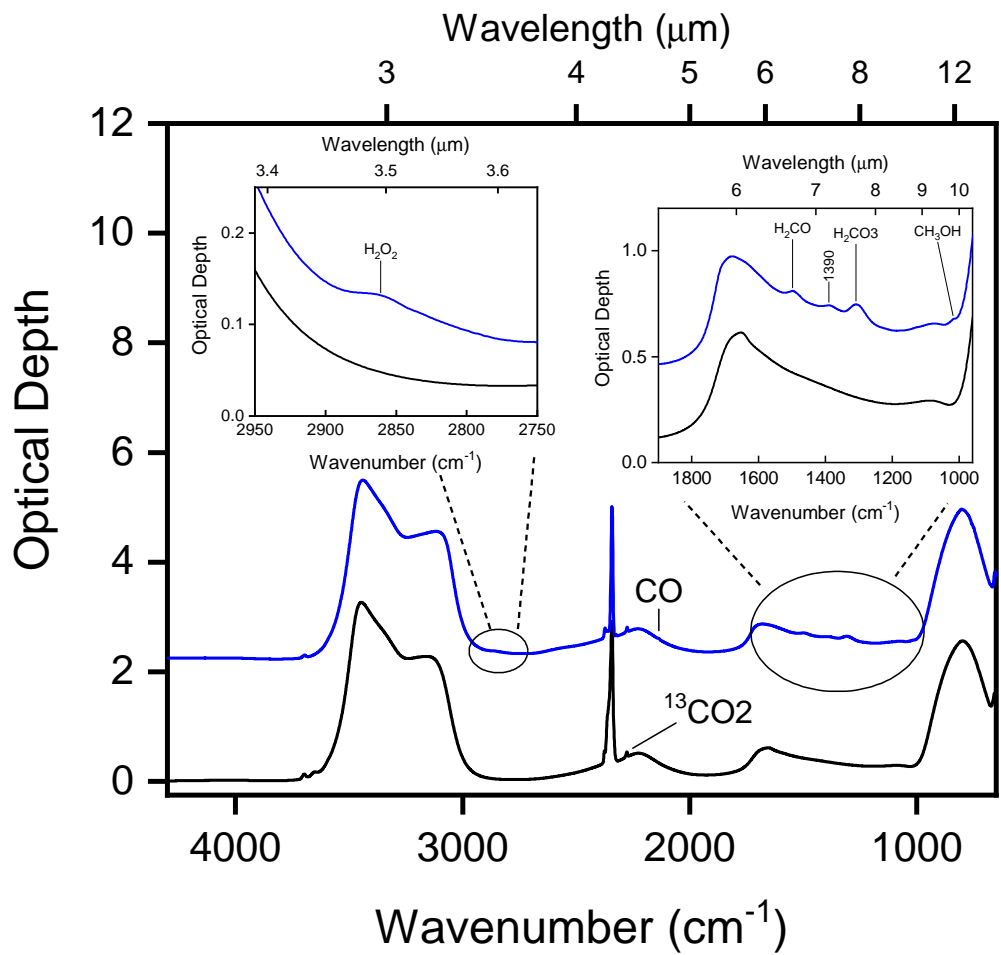

**Figure 5.** Spectra of a $\sim$ 2.76 µm thick $H_2O$ + 4.4% $CO_2$ ice before (bottom, black curve) and after (top, blue curve) irradiation to a fluence of $3.1 \times 10^{16}$ e$^-$ cm$^{-2}$. The left inset highlights the emergence of the $H_2O_2$ feature at 2861 cm$^{-1}$, analyzed as described in Section 2 and Figure 1. The right inset shows an expanded view of the mid-IR region, illustrating the formation of new CHO-organics absorptions at 1498, 1390, 1307, and 1016 cm$^{-1}$ after irradiation. The spectra have bee offset vertically for clarity.

## C. ISOLATING $H_2O_2$ FROM CHO ORGANICS BY SPECTRAL DECONVOLUTION

To estimate the 'organics' contribution to the 3.5 µm band, we fit the irradiated spectra as a sum of an $H_2O_2$ component and three Gaussian peaks representing overlapping CHO-organic absorptions. The $H_2O_2$ component, derived from the pure water ice irradiation experiment, was scaled and redshifted ($< 7$ cm$^{-1}$) to match the intensity and peak position of the 3.5 µm band for the mixed ice. One Gaussian, centered near 2831 cm$^{-1}$ (3.53 µm), likely captures the methanol contribution, while the other two represent additional C-H stretches likely from formaldehyde [$H_2CO$; (Schutte et al. 1993)] or O-H stretch from carbonic acid [$H_2CO_3$; (Brucato et al. 1997; Gerakines et al. 2000)]. Adding these CHO components reduces the residual between the measured spectra and modeled spectra (Figure 6B). From this spectral deconvolution, we find that CHO-organics contribute $< 5\%$ to the 3.5 µm band area in ice film with 2.5% $CO_2$. In ices containing 8.7% $CO_2$ (maximum $CO_2$ inclusion), the organic contribution to the band area is below 15%.

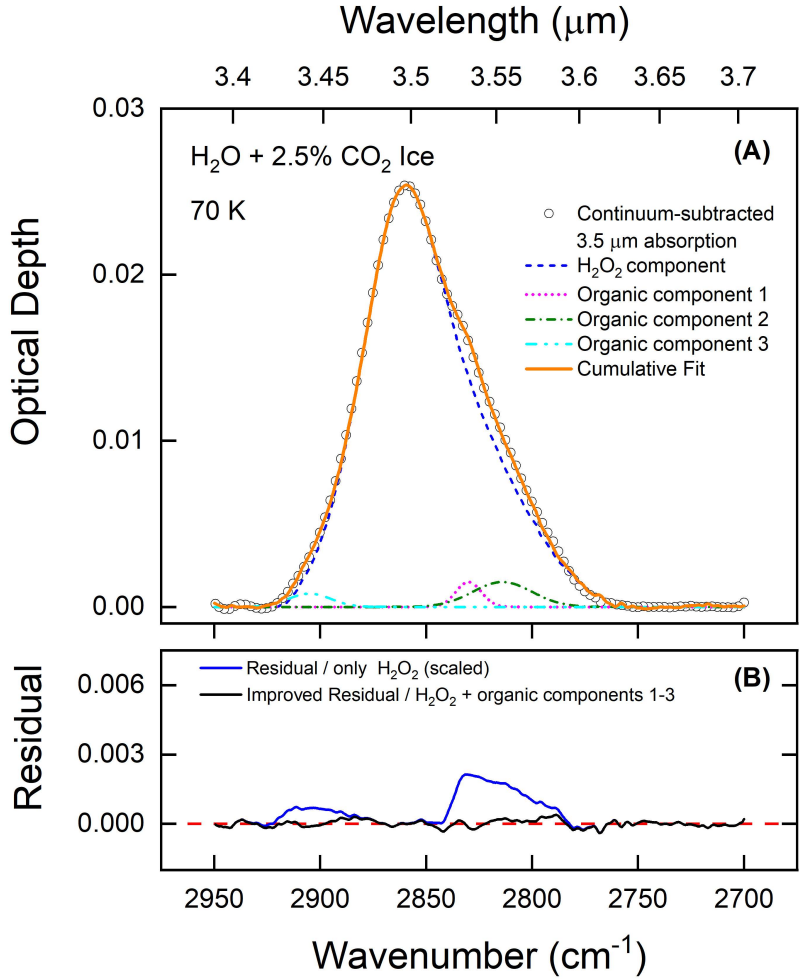

**Figure 6.** **A**: Spectral deconvolution of the continuum-subtracted 3.5 µm absorption (black open circles) obtained following 10 keV electron irradiation of water ice with 2.5% (by number) $CO_2$ at 70 K. The absorption is modeled using a scaled $H_2O_2$ component (blue dashed line) from pure $H_2O$ ice irradiation, combined with three Gaussian components (magenta, olive, and cyan dashed lines) representing CHO-organics that have overlapping absorptions with $H_2O_2$. The first organic component, centered at 2831 cm$^{-1}$ (3.53 µm, magenta dashed line), is methanol. Although specific assignments for the other two Gaussian components are uncertain, C-H stretches from other organics likely contribute to these features, adding a broadened shoulder on the long-wavelength side of the peroxide absorption. The cumulative fit (solid orange line) shows good agreement with the measured spectrum. **B**: Improvement in residuals ($Data - Fit$) when organic components are included. The blue residual curve represents $H_2O_2$ only, while the black curve includes CHO species. The $H_2O_2$ column densities in $CO_2$-containing ices are calculated after organic contributions ($< 15\%$) have been subtracted from the total integrated band area.

## D. RADIOLYTIC DECOMPOSITION OF $CO_2$ AND THE FORMATION OF CHO-ORGANICS

We track $CO_2$ loss with irradiation fluence by monitoring the baseline-subtracted area of the 4.27 μm $\nu_3$ band, which is not saturated (i.e. reflectance at peak absorption wavelength $> 0$) in films with $CO_2$ content $< 1\%$. Following irradiation to the highest dose, $\sim 75\%$ of the initial $CO_2$ remains intact (Figure 7A).

Quartz crystal microbalance (QCM) desorption profiles (Figure 7A, inset) show that $> 60\%$ of the consumed $CO_2$ re-emerges as carbonic acid ($H_2CO_3$); the remainder forms more volatile CHO-products such as methanol ($CH_3OH$) and formaldehyde ($H_2CO$) (Figure 5) that desorb before or with $H_2O$ at $\sim 180$ K. The formation of $H_2CO_3$ is further confirmed by infrared spectroscopy, exhibiting prominent characteristic absorption peaks at 1301, 1506, 1724, and 2621 cm$^{-1}$ (Figure 7B). The stronger QCM desorption trace at 100 K than at 70 K (Figure 7A, inset) indicates that warmer ices favor the synthesis of $H_2CO_3$.

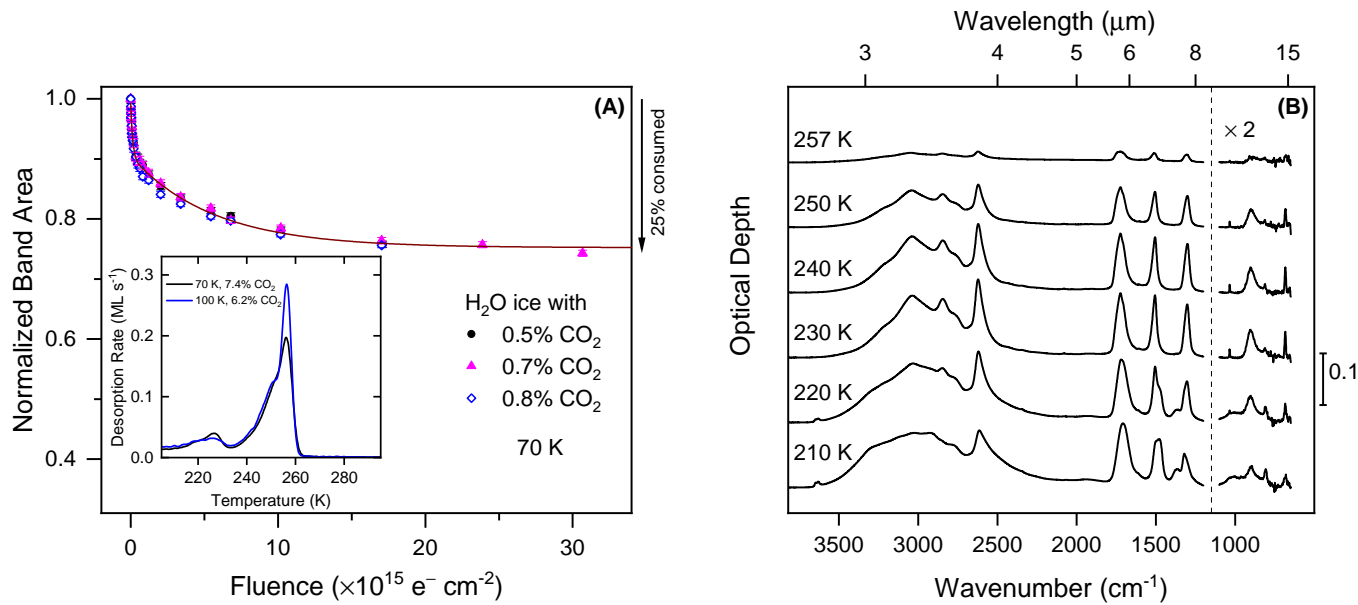

**Figure 7. A:** Normalized band area of the $CO_2$ 4.27 μm absorption band vs. fluence during 10 keV electron irradiation of $H_2O$ ice with varying initial $CO_2$ concentration. The films were irradiated at 70 K until $H_2O_2$ saturation was reached, at which point $\sim 25\%$ of the $CO_2$ was consumed. **A, inset:** Desorption rate (in equivalent $H_2CO_3$ ML s$^{-1}$) vs. temperature of radiolytic C-bearing residue formed after irradiating water ice film containing 7.4% CO2 at 70 K (black curve). Most ($> 60\%$) of the consumed $CO_2$ contributes to the formation of $H_2CO_3$, which desorbs near 255 K. The blue curve in the inset represents a film with 6.2% $CO_2$ irradiated at 100 K, showing organic production is enhanced by 14% compared to the colder 7.4% $CO_2$ film. **B:** Comparison of the continuum-subtracted IR spectra of the organics residue at the indicated temperatures for the film with 7.4 % $CO_2$. This thin film fully desorbs above 260 K. The spectra have been offset vertically for clarity.

## E. DESTRUCTION CROSS SECTION: EMPIRICAL MODEL FIT

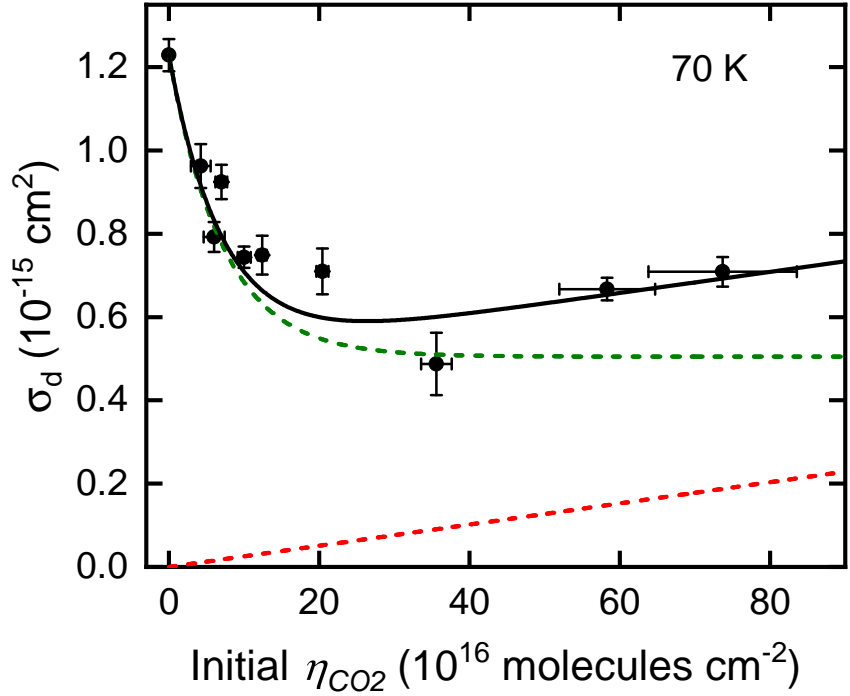

**Figure 8.** Fit to the 70 K destruction cross section values (solid black curve) using the empirical relation $\sigma_d = \sigma_{d_0} - \sigma_{s_1}(1 - \exp(-\sigma_{s_0}\eta_{CO_2})) + \sigma_{d_1}(\eta_{CO_2}/\eta_T)$. The cross section decreases sharply with small amounts of $CO_2$, as $CO_2$ or its radiolytic by-products protect $H_2O_2$ from destruction by reacting with OH radicals to form organics, or by scavenging secondary electrons. This shielding effect (olive dashed line) is represented by the first two terms in the empirical equation. At higher $CO_2$ abundances, the shielding effect diminishes, and the third term, representing $CO_2$-mediated $H_2O_2$ destruction (red dashed line), becomes more significant. The parameters values used to obtain the fit are $\sigma_{s_0} = (1.4 \pm 0.2) \times 10^{-17}$ cm$^2$, $\sigma_{s_1} = (7.3 \pm 0.1) \times 10^{-16}$ cm$^2$, and $\sigma_{d_1} = (2.1 \pm 0.2) \times 10^{-15}$ cm$^2$. The corresponding values for the 100 K data are $\sigma_{s_0} = (1.9 \pm 0.2) \times 10^{-17}$ cm$^2$, $\sigma_{s_1} = (9.0 \pm 0.1) \times 10^{-15}$ cm$^2$, and $\sigma_{d_1} = (7.5 \pm 0.5) \times 10^{-15}$ cm$^2$. The parameter $\sigma_{d_0}$ is the cross section for pure water ice.